\documentclass{elsart5p}

\usepackage{graphics}
\usepackage{graphicx}
\usepackage{amssymb}
\begin{document}

\begin{frontmatter}

\title{Crystal growth and anisotropic magnetic properties of RAg$_2$Ge$_2$\\ (R = Pr, Nd and Sm) single crystals.}

\author{Devang A. Joshi}, \author{R. Nagalakshmi}, \author{R. Kulkarni}, \author{S. K. Dhar}, and \author{A. Thamizhavel\corauthref{Thamizhavel}} \ead{thamizh@tifr.res.in},

\address{Department of Condensed Matter Physics and Material Sciences, \\Tata Institute of Fundamental Research, Dr. Homi Bhabha Road, Colaba, Mumbai 400 005, India}

\corauth[Thamizhavel]{Corresponding author. Tel/Fax: +91 22 22782438}

\begin{abstract}

We report the single crystal growth and anisotropic magnetic properties of the tetragonal RAg$_2$Ge$_2$ (R = Pr, Nd and Sm) compounds which crystallize in the ThCr$_2$Si$_2$ type crystal structure with the space group \textit{I4/mmm}.  The single crystals of RAg$_2$Ge$_2$ (R = Pr, Nd and Sm) were grown by self-flux method using Ag:Ge binary alloy as flux. From the magnetic studies on single crystalline samples we have found that PrAg$_2$Ge$_2$ and NdAg$_2$Ge$_2$ order antiferromagnetically at 12~K and 2~K respectively, thus corroborating the earlier polycrystalline results. SmAg$_2$Ge$_2$ also orders antiferromagnetically at 9.2~K.  The magnetic susceptibility and magnetization show a large anisotropy and the easy axis of magnetization for PrAg$_2$Ge$_2$ and NdAg$_2$Ge$_2$ is along the [100] direction where as it changes to [001] direction for SmAg$_2$Ge$_2$. Two metamagnetic transitions were observed in NdAg$_2$Ge$_2$ at $H_{\rm m1}$~=~1.25~T and $H_{\rm m2}$~=3.56~T for the field parallel to [100] direction where as the magnetization along [001] direction was linear indicating the hard axis of magnetization.
\end{abstract}


\end{frontmatter}

\section{Introduction}

RT$_2$X$_2$ (R: rare earths, T: Transition metal and X: p-block element (Si or Ge)) family of compounds crystallizing in the ThCr$_2$Si$_2$-type crystal structure with the space group $\textit{I4/mmm}$ have been extensively studied owing to their interesting magnetic properties, heavy-fermion superconductivity, pressure-induced superconductivity, unconventional metamagnetic transition etc.  Although there are many reports on the silicides and germanides of the 1-2-2 type structure, RAg$_2$Ge$_2$ compounds have not been investigated thoroughly.  Recently, we reported the anisotropic magnetic behaviour of CeAg$_2$Ge$_2$ single crystals~\cite{Thamizhavel}.   We have confirmed the antiferromagnetic ordering temperature $T_{\rm N}$ of this compound as 4.6~K, thus removing the uncertainty about $T_{\rm N}$ that existed in the literature.  Despite the fact that the Ce-ions in CeAg$_2$Ge$_2$ possess a tetragonal site symmetry, the ground state is a quartet, which was satisfactorily  explained by the heat capacity and the magnetic susceptibility data.  In continuation of our studies on CeAg$_2$Ge$_2$, we report in this paper on the crystal growth and anisotropic magnetic properties of RAg$_2$Ge$_2$ (R = Pr, Nd and Sm) series of compounds.  Earlier studies on polycrystalline samples revealed an antiferromagnetic ordering in  NdAg$_2$Ge$_2$ and in the case of PrAg$_2$Ge$_2$ an unconfirmed magnetic ordering was reported at 12~K~\cite{Szytula}. To further investigate these compounds for their anisotropic properties and to  confirm the polycrystalline results, we have succeeded in growing the single crystals of RAg$_2$Ge$_2$.

\section{Experimental Details}

The incongruent melting nature of RAg$_2$Ge$_2$  (R = Pr, Nd and Sm) precludes the single crystal growth from a direct melt.  Hence the RAg$_2$Ge$_2$ single crystals was grown by high temperature solution growth or the so called flux growth using a suitable solvent.  Since all the three constituent elements of RAg$_2$Ge$_2$  have high melting points none of them can be used as a flux.  The use of the fourth element as a flux may create the problem of forming a new phase.  From the binary phase diagram of Ag:Ge and found that for a particular compostion of Ag:Ge in the ratio 75.5 : 24.5 there is an eutectic point at 651~$^\circ$C.  We have taken advantage of this eutectic composition and used it as a flux for the growth of RAg$_2$Ge$_2$ single crystals.  The starting materials with 3N-Ce, 5N-Ag and 5N-Ge were taken in the ratio 1 : 16.25 : 6.75 and placed in a high quality recrystallized alumina crucible.  The alumina crucible was then subsequently sealed inside an evacuated quartz ampoule with a partial pressure of argon gas.  The temperature of the furnace was raised to 1050~$^\circ$C and held at this temperature for 48 hours in order to achieve proper homogenization.  Then, the temperature of the furnace was cooled down to the eutectic temperature of the binary flux Ag:Ge over a period of 3 weeks time and then rapidly cooled down to room temperature.  The crystals which are now embedded in the solidified flux matrix were then removed by means of centrifuging which was done at 750~$^\circ$C.  The crystals were mostly platelet like and the typical size of the crystals obtained was 5~$\times$~3~$\times$~0.4~mm$^3$.  The flat plane of the platelets corresponded to the (001) face of the crystal.  The crystals were then oriented along the principal crystallographic directions by means of Laue diffraction.  Well defined Laue diffraction spots, corresponding to  tetragonal symmetry pattern, indicated the good quality of single crystals.  The crystals were then cut along the crystallographic directions using spark erosion cutting machine for the anisotropic magnetic measurements.  The DC magnetic susceptibility and the magnetization measurements were performed in the temperature range 1.8 - 300~K and in  magnetic fields up to 7~T along the two principal directions using a Quantum Design SQUID magnetometer; high field magnetization measurements up to a field of 12~T were performed using a vibrating samle magnetomter (VSM, Oxford Instruments).

\section{Results and Discussion}

\begin{figure}[!]
\includegraphics[width=0.45\textwidth]{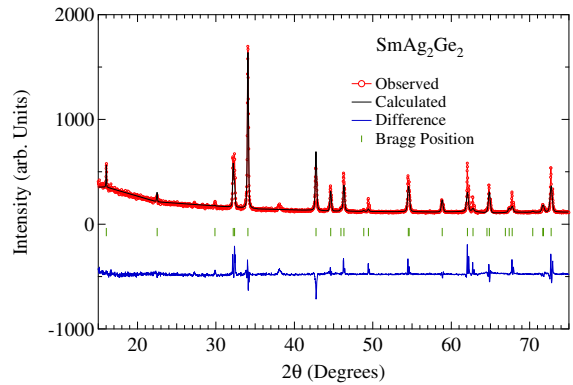}
\caption{A representative x-ray powder diffraction pattern of SmAg$_2$Ge$_2$.  The solid line through the experimental data points is the Rietveld refinement profile fitting for tetragonal SmAg$_2$Ge$_2$.  The Bragg positions and the difference between the observed the calculated patterns are also shown.} \label{fig1}
\end{figure}

We have performed the powder x-ray diffraction, by crushing a few pieces of the single crystals to confirm the phase purity, as the crystals were grown from a very off-stoichiometric melt.  Figure~\ref{fig1} shows the representative powder x-ray diffraction for SmAg$_2$Ge$_2$ together with the Rietveld refinement.  As  can be seen from the figure the powder x-ray diffraction clearly reveals that the grown single crystals are single phase with no detectable traces of impurity phase.  From the Rietveld refinement the ThCr$_2$Si$_2$-type crystal structure of SmAg$_2$Ge$_2$ is confirmed with the rare earth occupying the $2a$ site, the silver atom at the $4d$ site, while the Ge atom occupies the $4e$ site with the $z$ parameter 0.61240.  The lattice constants thus estimated for SmAg$_2$Ge$_2$ $a$~=~4.226(2)~\AA~and $c$~=~11.045(8)~\AA~with the reliability parameters $R_{\rm B}$~=~10\% and $R_{\rm F}$~=~6.03~\%.  The lattice constants of PrAg$_2$Ge$_2$ and NdAg$_2$Ge$_2$ match well with the previous reported values.  Furthermore, the energy dispersive X-ray analysis (EDAX) on all the grown single crystals confirmed the 1-2-2 stoichiometry.  

\begin{figure}[!ht]
\includegraphics[width=0.5\textwidth]{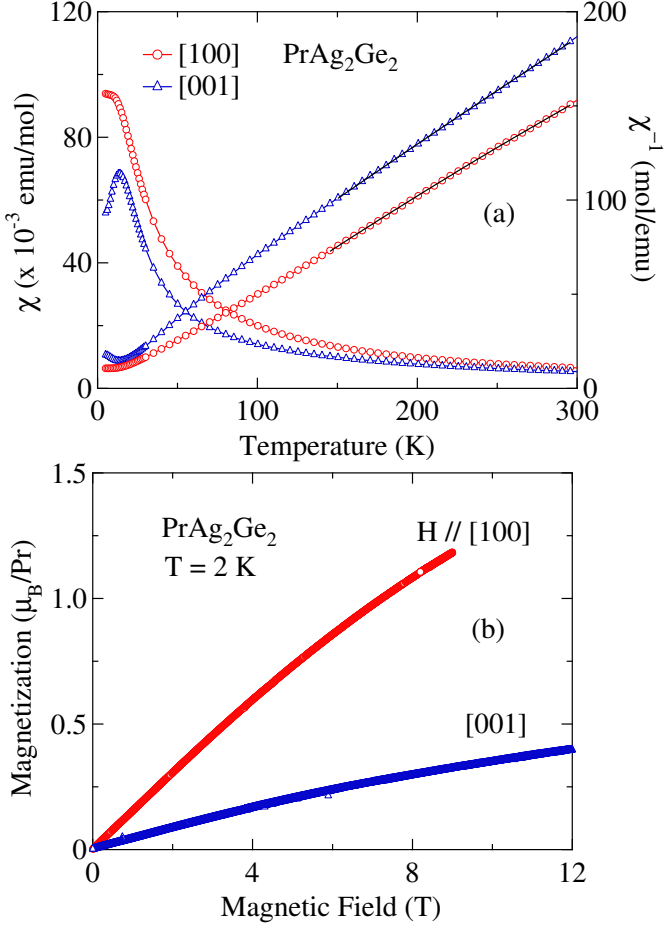}
\caption{(a)Temperature dependence of the magnetic susceptibility ($\chi$ and  $\chi^{-1}$) of PrAg$_2$Ge$_2$ in the range 1.8 - 300 K, the inset shows the low temperature part and (b) isothermal magnetization of PrAg$_2$Ge$_2$ at T = 2 K} \label{fig2}
\end{figure}

In Fig.~\ref{fig2}(a) we present the results of the anisotropic dc magnetic susceptibility $\chi$(T) and $\chi^{-1}$ for PrAg$_2$Ge$_2$ single crystal in an applied magnetic field of 0.1~T in the temperature range 1.8 - 300~K.  The susceptibility shows a large anisotropy for the field along [100] and [001] directions.  At high temperatures the susceptibility exhibits a Curie-Weiss like behavior and at low temperatures, the susceptibility along the [100] direction increases more rapidly than along the [001] direction indicating that [100] is th easy axis of magnetization.  It is to be noted that at $T_{\rm N}$~=~12~K there is a change of slope in the magnetic susceptibility for $H~\parallel$~[100], while for $H~\parallel$~[001], the susceptibility shows a sudden drop typical for an antiferromagnet.  This indicates that PrAg$_2$Ge$_2$ exhibits a complex magnetic behaviour just below $T_{\rm N}$ at 12~K. 

The high temperature magnetic susceptibility along both the principal directions was fitted to a modified Curie-Weiss expression,
\begin{center}
\begin{equation} 
\label{eqn1}
\chi = \chi_0 + \frac{C}{(T-\theta_{\rm p})},
\end{equation}
\end{center}

where $\chi_0$ is the temperature independent magnetic susceptibility which includes the core-electron diamagnetism and the susceptibility of the conduction electrons, $C$ is the Curie constant which can be expressed in terms of the effective moment as
\begin{equation}
\label{eqn2}
C = \frac{\mu_{\rm eff}^2~x}{8},
\end{equation}
where $x$ is the number of rare-earth atoms per formula unit.  The solid lines in Fig.~\ref{fig2}(a) are fits to the Eqn.~\ref{eqn1} with the fitting parameters estimated as $\chi_0$ = 8~$\times~10^{-4}$, $\mu_{\rm eff}$~=~3.58~$\mu_{\rm B}$/Pr and $\theta_{\rm p}$~=~22~K for field parallel to [100] direction and $\chi_0$ = 3.1~$\times~10^{-4}$, $\mu_{\rm eff}$~=~3.58~$\mu_{\rm B}$/Pr and $\theta_{\rm p}$~=~-16~K for field parallel to [001] direction.  The value of the effective moment from the fitting procedure shows that the Pr-ions are in a trivalent state.

The isothermal magnetization of PrAg$_2$Ge$_2$ at $T$~=~2~K is shown in Fig.~\ref{fig2}(b).  The magnetization increases almost linearly along both the principal directions.  The magnetization for $H~\parallel$~[100] is higher compared to that of [001] corroborating the [100] direction as the easy-axis of magnetization in PrAg$_2$Ge$_2$.  The magnetization along [100] direction amounts to 1.17~$\mu_{\rm B}$/Pr at 8~T and for field along [001] direction the magnetization amounts to 0.39~$\mu_{\rm B}$/Pr at 12~T.  These values are smaller compared to the saturation moment ($g_JJ = 3.2$~$\mu_{\rm B}$) which indicates that a high magnetic field is necessary to reach the saturation moment of Pr.

\begin{figure}[!ht]
\includegraphics[width=0.5\textwidth]{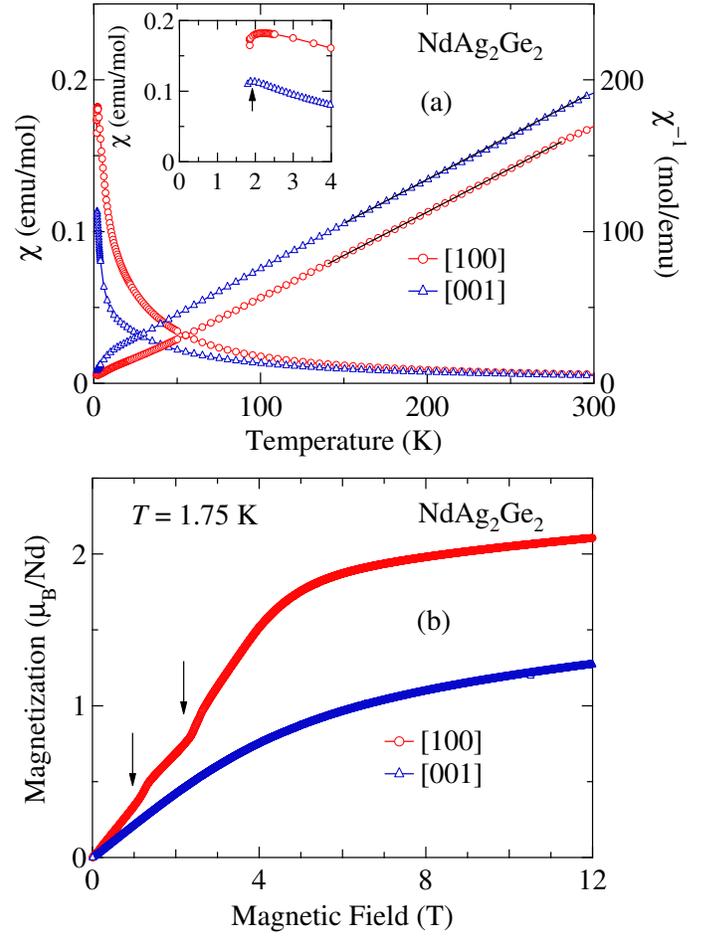}
\caption{(a) Temperature dependence of the magnetic susceptibility ($\chi$ and $\chi^{-1}$ of NdAg$_2$Ge$_2$ in the range 1.8 - 300 K, (b) isothermal magnetization of NdAg$_2$Ge$_2$ at T = 1.75 K} \label{fig3}
\end{figure}

Figure~\ref{fig3}(a) shows the temperature dependence of magnetic susceptibility of NdAg$_2$Ge$_2$ along the two principal directions, measured in an applied field of 0.5~T.  Similar to PrAg$_2$Ge$_2$, the susceptibility along the [100] direction is larger compared to that of [001] direction indicating that the easy axis of magnetization is [100] direction.  The drop in the magnetic susceptibility at 2~K indicate the magnetic ordering of NdAg$_2$Ge$_2$ at 2~K.  The magnetic ordering is confirmed to be antiferromagnetic in nature from the results of isothermal magnetization, as explained later.  The $T_{\rm N}$ at 2~K is indicated by the arrows in the inset of Fig.~\ref{fig3}(a). The inverse magnetic susceptibility of NdAg$_2$Ge$_2$ is also shown in Fig.~\ref{fig3}(a).  The solid lines are fits to the Eqn.~\ref{eqn1}. The obtained fitting parameters are $\chi_0$~=~1.9~$\times$~10$^{-3}$, $\mu_{\rm eff}$~=~3.62~$\mu_{\rm B}$/Nd and $\theta_{\rm p}$~=~-25~K for $H~\parallel$~[001] and $\chi_0$~=~2.5~$\times$~10$^{-3}$, $\mu_{\rm eff}$~=~3.62~$\mu_{\rm B}$/Nd and $\theta_{\rm p}$~=~6~K for $H~\parallel$~[100].  The estimated effective magnetic moment clearly reveals that the Nd ions are trivalent.

The isothermal magentization of NdAg$_2$Ge$_2$ at $T$~=~1.75~K just below the antiferromagnetic ordering temperature exhibits  two metamagnetic transitions at $H_{m1}$~=~1.25~T and $H_{m2}$~=~3.56~T along the [100] direction and the saturation magnetic moment is found to be 2.1~$\mu_{\rm B}$/Nd at 12~T field.  These results of the magnetic moment clearly corroborates the previous neutron diffraction measurement made on a polycrystalline sample~\cite{Szytula}, in which it has been reported that the moments are aligned along the [100] direction with an ordered moment of 1.3(9)~$\mu_{\rm B}$/Nd.  The magnetization along [001] direction did not show any anomaly up to a field of 12~T where it attains a value of 1.26~$\mu_{\rm B}$/Nd indicating the hard axis of magnetization.  

\begin{figure}[!ht]
\includegraphics[width=0.5\textwidth]{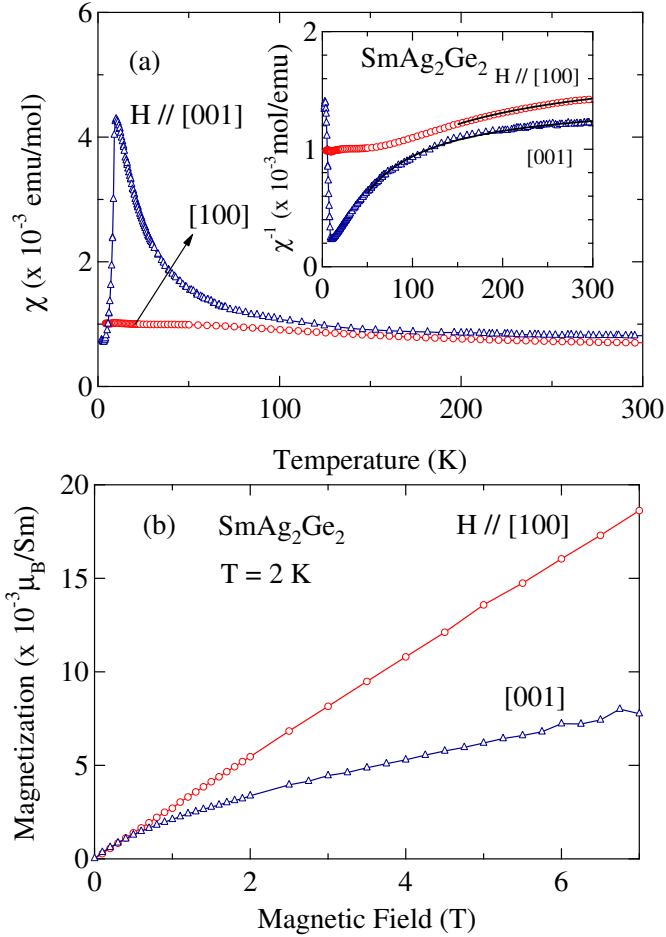}
\caption{(a)Temperature dependence of the magnetic susceptibility of SmAg$_2$Ge$_2$ in the range 1.8 - 300 K, the inset shows the inverse magnetic susceptibility (b) isothermal magnetization of SmAg$_2$Ge$_2$ at T = 2~K} \label{fig4}
\end{figure}

The temperature dependence of the magnetic susceptibility of SmAg$_2$Ge$_2$ is shown in Fig.~\ref{fig4}(a).  The antiferromagnetic ordering is clearly seen by the rapid drop of susceptibility at $T_{\rm N}$~=~9.2~K along the [001] direction.   The inset in Fig~\ref{fig4}(a) shows the inverse magnetic susceptibility of SmAg$_2$Ge$_2$.  The magnetic susceptibility of Sm compounds without considering the crystalline electric field (CEF) effect can be expressed as,
\begin{equation}
\label{eqn3}
\chi = \frac{N_A~\mu_{\rm B}^2}{k_B}\left(\frac{20}{7~\Delta E} + \frac{\mu_{\rm eff}^2}{3~(T - \Theta)}   \right),
\end{equation}
where $\Delta$E is the average energy between the $J$~=~5/2 ground state multiplet and the $J$~=7/2 excited state multiplet.  The solid line in the inset of Fig.~\ref{fig4}(a) is the fit to Eqn.~\ref{eqn3}.  The fitting parameters were estimated to be $\mu_{\rm eff}$~=~0.49~$\mu_{\rm B}$/Sm, $\Theta$~=~15~K and $\Delta$E~=~1537~K for $H~\parallel$~[001] and $\mu_{\rm eff}$~=~0.58~$\mu_{\rm B}$/Sm, $\Theta$~=~-14~K and $\Delta$E~=~1903~K $H~\parallel$~[100], the values of $\Delta$E are close to the value of $\Delta$~$\simeq$~1500~K estimated for a free Sm$^{3+}$ ion.  

The isothermal magnetization of SmAg$_2$Ge$_2$ measured at $T$~=~2~K is shown in Fig.~\ref{fig4}(b).  The magnetization measurements were performed up to a field of 6~T and  no anomaly is seen in both the principal directions.  It is to be noted here that the magnetization along the hard axis namely, the [100] direction is higher than that along the  [001] direction. Similar kind of anomalous magnetization behavior is observed in SmAgSb$_2$~\cite{Myers}. The smaller value of the magnetization along the easy axis, means that there is a possibility of metamagnetic transition along [001] direction like a spin-flop or spin-flip type, at high magnetic fields, where the magnetization will saturate much faster than that along the [100] (hard) direction.  High field magnetization measurement is necessary to confirm this.  The saturation moment at 6~T amounts to 0.018~$\mu_{\rm B}$/Sm for $H~\parallel$~[100] and 0.008~$\mu_{\rm B}$/Sm which is smaller than the saturation moment (0.71~$\mu_{\rm B}$) of Sm ions.

\section{Summary and Conclusion}
We have successfully grown the single crystals of RAg$_2$Ge$_2$ (R = Pr, Nd and Sm) and studied the anisotropic magnetic properties by magnetic susceptibility and magnetization measurements.  All the three compounds order antiferromagnetically with the easy axis of magnetization along [100] direction for PrAg$_2$Ge$_2$ and NdAg$_2$Ge$_2$ where as the easy axis of magnetization changes to [001] direction for SmAg$_2$Ge$_2$. Our single crystal data on PrAg$_2$Ge$_2$ clearly indicates the antiferromagnetic ordering which was not discernible from  neutron scattering experiment on the polycrystalline samples. It is to be noted here that the N\'{e}el temperature of PrAg$_2$Ge$_2$ ($T_{\rm N}$~=~12~K) is higher than that of NdAg$_2$Ge$_2$ ($T_{\rm N}$~=~2~K), furthermore, the $T_{\rm N}$ of PrAg$_2$Ge$_2$ is close to that of SmAg$_2$Ge$_2$ ($T_{\rm N}$~=~10~K) which means that RAg$_2$Ge$_2$ do not obey the usual de Gennes scaling.   If the magnetic ordering is mainly due to RKKY interaction, then the magnetic ordering temperature can be scaled by the well known de Gennes scaling factor : $(g-1)^2J(J+1)$.  The higher ordering temperature in PrAg$_2$Ge$_2$ may presumably be attributed to the crystal field effects. This kind of deviation has been observed in the RT$_2$X$_2$ (X = Si or Ge) series of compounds~\cite{Szytula2}.

\section{Acknowledgments}

One of the authors A. Thamizhavel would like to acknowledge the SCES-2008 organizers and the ICAM-I2CAM for the financial support to attend the International Symposium on Strongly Correlated Electron System (SCES)- 2008.

\end{document}